\documentclass{iopjournal}

\usepackage{amsmath,amssymb}
\usepackage{graphicx}
\usepackage{lmodern}
\usepackage{caption}
\usepackage[%
backend=biber,%
style=phys,%
autocite=plain,%
articletitle=false,biblabel=brackets,%
chaptertitle=false,pageranges=false%
]{biblatex}
\begin{document}

\articletype{Paper} 

\title{Memory-controlled random bit generator}

\author{Mateusz Wi\'{s}niewski$^1$\orcid{0000-0002-0396-1427} and Jakub Spiechowicz$^{1,*}$\orcid{0000-0001-7569-4826}}

\affil{$^1$Institute of Physics, University of Silesia in Katowice, Chorzów, Poland}

\affil{$^*$Author to whom any correspondence should be addressed.}

\email{jakub.spiechowicz@us.edu.pl}


\begin{abstract}
Nowadays a bit is no longer a mere abstraction but a physical quantity whose manipulation governs both operation of modern technologies and theoretical frontiers of fundamental science.
In this work we propose a setup in which the memory time can be utilized to control the generation and storage of binary information. In particular, we consider a nonequilibrium Brownian particle immersed in a viscoelastic environment and dwelling in a spatially periodic potential. We interpret its average velocity as a bit and show that depending on the memory time characterizing the viscoelastic bath the particle can be either in one of two stable states representing the bit values or in a chaotic state in which the information is erased and a new bit can be generated. We analyze randomness of the so obtained bit sequence and assess the stability of the produced values. Our study provides a blueprint for storing and processing information in a microscopic system using its memory.
\end{abstract}

\section{Introduction}

The notion of information, measured in a fundamental unit of a bit, has become indispensable for modern physics, linking the abstract domain of communication theory with the core principles of statistical mechanics, thermodynamics, and quantum theory. Initially conceived by Shannon \cite{Shannon1948} to formalize the efficiency of communication channels, information theory now provides a rigorous language to quantify entropy, correlations, and complexity in physical systems \cite{Vedral2002, Landi2021}. At the quantum scale, bits generalize to qubits, and the interplay between entanglement, measurement, and information processing defines new paradigms of computation and cryptography \cite{Galindo2002}. The recognition that "\emph{information is physical}" as manifested in Landauer's principle \cite{Landauer1991} has completely reshaped our understanding of this concept. Nowadays it is no longer a mere abstraction but exists only as a property (state) of a physical system and as such it is subjected to the fundamental laws of physics. Information processing governs both the operation of modern technologies and the theoretical frontiers of basic sciences.

On the other hand, in recent years the research on soft matter has become one of the hottest topics in physics and beyond. It has been shown that its viscoelastic behaviour has a profound impact on the dynamics of Brownian particles immersed in it, and can lead to such effects as subdiffusion \cite{Goychuk2012}, acceleration or slowdown of barrier crossing \cite{Kappler2018, Ferrer2021, Ginot2022}, circular motion of achiral microswimmers \cite{Narinder2018}, and induction of Magnus effect \cite{Cao2023}, negative mobility \cite{Wisniewski2024-nm} or current reversal \cite{Wisniewski2025-cr}, to name only a few. The importance of these discoveries stems from the fact that viscoelasticity is a property of many microbiological environments, such as cytoplasm \cite{Pullarkat2007} or blood \cite{Thurston1972}, but it is also observed in polymer networks \cite{Young2011}, micellar solutions \cite{Cates1990}, and liquid crystals \cite{Waigh2016}.

The viscoelastic properties of soft matter depend on its composition and can be tuned by changing the proportions of its constituents responsible for its elastic and viscous response \cite{Charrier2018}. 
So far, this method has been used to study the functioning of cells in extracellular environments with different mechanical properties \cite{Chaudhuri2015, Uto2016, Charrier2018, Tang2018}. In such a setting, however, the change of the viscoelastic properties of the setup requires the preparation of a new system, which does not allow for studying the effects of the temporal variations of the properties. The solutions to this problem are systems in which the internal cross-links can be rearranged after applying an external stimulus, such as light \cite{Marozas2019, Carberry2020, Lu2025} or an electric field \cite{Teng2024}. This method allows for the change of setup properties ``on the fly'' and makes its characteristic quantities, such as fluidity or memory time, new control parameters.

The ability to change properties of the environment on demand opens the possibility of controlling the dynamics of microscopic objects immersed in it by an external stimulus. In this article, we propose a setup in which varying viscoelasticity can be utilized to control the generation and storage of binary information. In particular, we consider a Brownian particle dwelling in a periodic potential and coupled to correlated thermal bath, and show that depending on the viscoelastic properties of the bath, the particle can be either in one of two stable states representing values of the information bit, or can be in an aperiodic state where the stored information is forgotten and a new bit value is generated. We analyze the stability of stored information and evaluate how random the generated bit sequence is. Our results advance the understanding of the role of the system environment in computation and information processing in microscopic systems.

\section{Model and methods}

Our system of interest consists of a Brownian particle of mass $m$ dwelling in a spatially periodic potential $U(x) = \Delta U\cos(2\pi x)$ and driven by an oscillatory force $F(t) = a\cos(\omega t)$. The temporal evolution of its position $x$ and velocity $v$ can be described by the Generalized Langevin Equation, reading \cite{Luczka2005}
\begin{equation} \label{eq:gle}
    m\dot{v} = -\gamma\int_0^t K(t-s)v(s)\mathrm{d}s - \frac{\mathrm{d}U(x)}{\mathrm{d}x} + F(t) + \eta(t).
\end{equation}
Eq.~\ref{eq:gle} can be recast to a dimensionless form in which the Stokes friction coefficient $\gamma \equiv 1$ and half of the barrier height of the periodic potential $\Delta U \equiv 1$ (see e.g.~Ref.~\cite{Wisniewski2022} for details on the appropriate length and time scales). Consequently, in the following we set $\gamma = \Delta U = 1$ and treat all other parameters as dimensionless quantities.

In this approach viscoelasticity of the surrounding medium is characterized by the memory kernel $K(t)$, which captures its response to external perturbations and determines the friction experienced by the Brownian particle. As dictated by the fluctuation-dissipation theorem \cite{Kubo1966}, the memory kernel is also related to the autocorrelation function of thermal fluctuations $\eta(t)$, namely 
\begin{equation}
    \langle \eta(t) \eta(s) \rangle = \gamma \theta K(|t-s|),
\end{equation}
where $\theta$ is the dimensionless temperature. 
In this article we assume that the memory kernel decays exponentially, i.e.
\begin{equation} \label{eq:K}
    K(t) = \frac{1}{\tau}e^{-t/\tau}.
\end{equation}
Such a form appears in Maxwell's model of viscoelasticity and it is characterized by a single characteristic time $\tau$, which can be interpreted as a memory time or correlation time of thermal fluctuations \cite{Goychuk2012}. In the limit $\tau\to0$, the kernel becomes $2\delta(t)$ and Eq.~\ref{eq:gle} simplifies to a memoryless Langevin Equation.

\subsection{Effective mass approach}
If the memory time $\tau$ is much shorter than the relaxation time of the free particle $\tau_L = m/\gamma$, Eq.~\ref{eq:gle} can be approximated with a memoryless Langevin equation \cite{Wisniewski2024-effmass,Wisniewski2024-emb}
\begin{equation} \label{eq:eff_mass}
    m^* \dot{v} = -\gamma v - \frac{\mathrm{d}U(x)}{\mathrm{d}x} + F(t) + \xi(t),
\end{equation}
where $m^* = m - \Delta m$ is the effective mass of the particle, and $\xi(t)$ is thermal white noise obeying $\langle \xi(t)\xi(s)\rangle = 2\gamma \theta\delta(t-s)$. The mass correction $\Delta m$ depends on the form of the memory kernel $K(t)$ and reads
\begin{equation} \label{eq:dm}
    \Delta m = \gamma\int_0^\infty tK(t) \mathrm{d}t.
\end{equation}
In the case of the exponentially decaying $K(t)$ (Eq.~\ref{eq:K}) the mass correction $\Delta m = \tau$ and the effective mass is simply
\begin{equation} \label{eq:m_star}
    m^* = m-\tau.
\end{equation}
Eq.~\ref{eq:eff_mass} is thus a bridge between the full description of the particle's dynamics given by the Generalized Langevin Equation (Eq.~\ref{eq:gle}) and its memoryless variant $\tau\to 0$. It allows for studying the influence of short memory on the dynamics of a Brownian particle with a memoryless equation and offers an appealing interpretation of the origin of memory-induced effects.

\subsection{Quantity of interest}
The presence of the periodic driving force $F(t)$ implies that the Brownian particle is not in equilibrium with thermal bath and that the dominating frequency in the power spectrum of its velocity is equal to $\omega$. To get rid of this periodic component, our main quantity of interest will be the velocity of the particle averaged over the period of the driving force $\mathsf{T} = 2\pi/\omega$
\begin{equation} \label{eq:v_av}
    \mathsf{v}(t) = \frac{1}{\mathsf{T}}\int_t^{t+\mathsf{T}} v(s)\mathrm{d}s.
\end{equation}
In the zero-temperature limit $\theta=0$ the particle in the asymptotic long time limit arrives at the dynamical attractor in the phase space and its instantaneous velocity $v(t)$ can be periodic 
quasiperiodic or chaotic \cite{Kautz1996}. Consequently, the period-averaged velocity $\mathsf{v}(t)$ can be constant, 
periodic with period $n\mathsf{T}$ 
or can exhibit no regularity if $v(t)$ is quasiperiodic or chaotic. The addition of thermal noise induces thermally activated escape events among coexisting attractors so that the period-averaged velocity fluctuates around a constant value or is not regular depending on whether thermal fluctuations perturb the deterministic regular or chaotic attractor.

\subsection{Methods of solution}
Eq.~\ref{eq:gle} is a nonlinear stochastic second-order integro-differential equation, and as such it cannot be solved analytically. 
In order to solve it numerically, we implemented a weak second-order predictor-corrector algorithm \cite{Platen2010} with a timestep $h=10^{-2}\times\mathsf{T}$. The particle trajectories were typically run for $10^{3}$ periods of the driving force $\mathsf{T}$ starting from different initial positions $x(0)$, velocities $v(0)$, and phases of the driving force $F(t)$. The numerical analysis was performed with the use of a Graphics Processing Unit (GPU) supercomputer, which allowed us to calculate the particle evolution for multiple initial conditions and realizations of the thermal noise in parallel \cite{Spiechowicz2015}.

\section{Results}
The goal of this paper is to present a setup in which, depending on the memory time characterizing thermal bath, the particle can be in one of two stable states representing bits of information, or in an irregular state where the information is lost and a new bit value is generated. Typically, the information is encoded in the position of the Brownian particle placed in a double-well potential, where each of the wells represents a stable state \cite{Parrondo2015}. Here we present another approach, in which the information is encoded in the period-averaged velocity $\mathsf{v}(t)$, which can be either positive or negative. We adopt the convention that the positive and negative state is identified with "1" and "0" bit, respectively. 
We set $m=1.0$, $a = 8$, $\omega = 5$, and $\theta=10^{-4}$ unless stated otherwise, but the principle of operation of our setup is rather general, as will be clarified later in this work. 
\begin{figure}[htbp]
    \centering
    \includegraphics{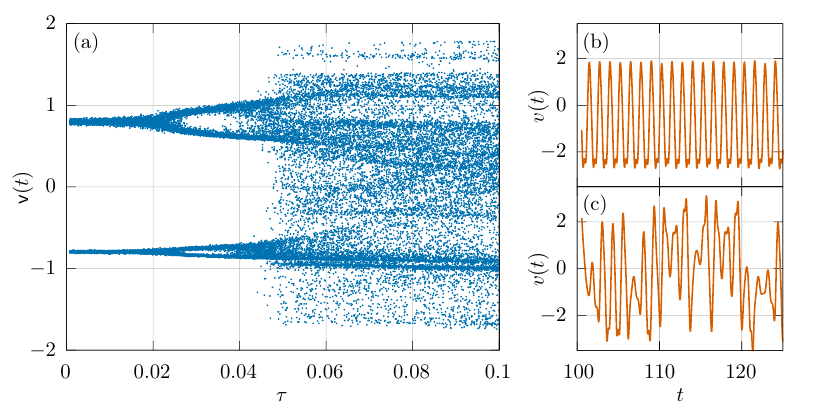}
    \caption{(a) "Bifurcation" diagram of the period-averaged velocity $\mathsf{v}(t)$ as a function of the memory time $\tau$. In (b) and (c) exemplary trajectories for $\tau=0.01$ and $\tau = 0.1$ are pictured.}
    \label{fig:bifurc_tau}
\end{figure}

We start our analysis with a "bifurcation" diagram of the period-averaged velocity $\mathsf{v}(t)$ as a function of the memory time $\tau$ presented in Fig.~\ref{fig:bifurc_tau}(a). The figure was obtained by solving Eq.~\ref{eq:gle} numerically for different initial conditions and realizations of the thermal noise $\eta(t)$ for $10^3$ periods of the driving force $\mathsf{T}$. Then the period-averaged velocity $\mathsf{v}(t)$ was calculated by averaging the instantaneous velocity $v(t)$ in each of the trajectories over the last period $\mathsf{T}$.

For memory times $\tau \lesssim 0.02$  (including the memoryless limit $\tau = 0$) the period-averaged velocity $\mathsf{v}(t)$ takes only two values $\mathsf{v}_{\pm} = \pm \omega/(2\pi) \approx \pm 0.8$ which are smeared due to to the presence of weak thermal noise. It is a consequence of $v(t)$ being periodic with period $\mathsf{T}$, see Fig.~\ref{fig:bifurc_tau}(b). The values $\mathsf{v}_\pm$ correspond to running solutions in which the particle travels one spatial period of the potential $U(x)$ during every period of the driving force $F(t)$ either in the positive or in the negative direction. The fact that there are no points between these may suggest that the particle rarely switches between these two solutions. The presence of two attractors with opposite period-averaged velocities is a consequence of the system's spatial symmetry. For this reason the average velocity of the setup must vanish identically. It implies that in the deterministic limit $\theta = 0$ every possible trajectory of the system is accompanied by the corresponding one propagating in the opposite direction \cite{Denisov2014}.  Consequently, if there is a deterministic attractor in which the period-averaged velocity equals $\mathsf{v}_+$, there also must be an attractor with $\mathsf{v}_- = -|\mathsf{v}_+|$.  

In contrast, when the memory time $\tau$ is longer, i.e. $\tau \gtrsim 0.05$, the period-averaged velocity $\mathsf{v}(t)$ takes values from almost the whole range between $-1.8$ and $1.8$. The reason is that in this parameter regime the instantaneous velocity $v(t)$ is aperiodic, see Fig.~\ref{fig:bifurc_tau}(c), and its period average can take different values depending on the initial conditions and a moment of time. This system thus meets our requirements. For $\tau < 0.02$ the particle can be in one of two stable states representing two values of an information bit. We assume that $\mathsf{v}_+$ and $\mathsf{v}_-$ renders the logic "1" and "0", respectively. For $\tau > 0.05$ the velocity exhibits no regularity and this state can be utilized as a generator of random bit values.
\begin{figure}[htbp]
    \centering
    \includegraphics{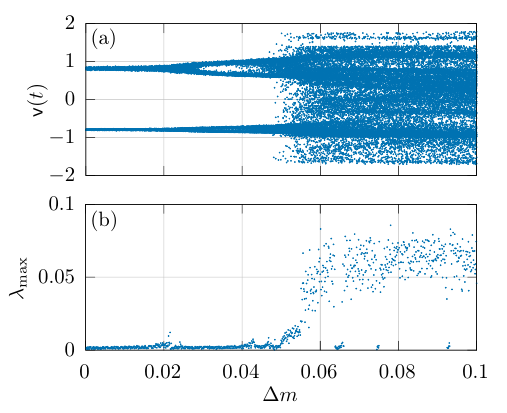}
    \caption{(a) "Bifurcation" diagram of the period-averaged velocity $\mathsf{v}$ for the approximate system in the effective mass approach as a function of the mass correction $\Delta m = \tau$. (b) Corresponding maximal Lyapunov exponent $\lambda_\mathrm{max}$ in the deterministic limit $\theta=0$ estimated using the method of reconstruction of the attractor based on the particle's trajectory \cite{Rosenstein1993}.}
    \label{fig:bifurc_m}
\end{figure}

Let us now find out why increasing the memory time $\tau$ results in the emergence of a qualitatively new aperiodic solution and the extinction of the periodic ones. According to the effective mass approach, the presence of short memory is approximately equivalent to a correction $\Delta m$ to the particle's mass in a memoryless system, see Eqs~\ref{eq:eff_mass}--\ref{eq:m_star}. Thanks to the simple formula for the mass correction $\Delta m = \tau$ obtained for the studied memory kernel (Eq.~\ref{eq:K}), the original dynamics studied as a function of the memory time $\tau$ is approximately equivalent to the memoryless dynamics studied as a function of $\Delta m$. In Fig.~\ref{fig:bifurc_m}(a) we plot the "bifurcation" diagram of the period-averaged velocity $\mathsf{v}(t)$ as a function of $\Delta m$, obtained by solving the approximate Eq.~\ref{eq:eff_mass}. Its stunning similarity to the diagram presented in Fig.~\ref{fig:bifurc_tau}(a) confirms that our setup is within the range of validity of the effective mass approach (i.e. $\tau \ll m/\gamma$). Moreover, it shows that the aperiodic solution arising upon increasing the memory time $\tau$ is also present in the memoryless dynamics for a lower mass of the particle $m^* = m - \Delta m$. 

To further quantify the dynamics of the studied setup, we estimate the maximum Lyapunov exponent in the deterministic limit of Eq.~\ref{eq:eff_mass}. For $\theta=0$ the system can be recast into a set of three autonomous equations for the phase variables $\mathcal{X}(t) = [x(t),\ v(t),\ \phi(t)=\omega t]$ reading
\begin{equation}
    \dot{\mathcal{X}}(t) = \mathcal{F}[\mathcal{X}(t)],
\end{equation}
where $\mathcal{F}[\mathcal{X}(t)] = [v(t),\ -\gamma v(t) - \frac{\mathrm{d}U(x(t))}{\mathrm{d}t} + F(t),\ \omega]$. If we now consider an infinitesimal ellipsoid in the phase space with the principal axes spanned along the phase space ones, the evolution of its volume $\mathcal{V}(t)$ can be expressed as \cite{Ott2002}
\begin{equation}
    \mathcal{V}(t) = \mathcal{V}(0) e^{(\lambda_x + \lambda_v + \lambda_\phi)t} = \mathcal{V}(0) e^{-\gamma t},
\end{equation}
where $\lambda_x$, $\lambda_v$ and $\lambda_\phi$ are the Lyapunov exponents corresponding to the phase variables. Since the system is dissipative, the sum of the Lyapunov exponents must be negative, and the volume of the initial ellipsoid decreases in time. The exponent $\lambda_\phi$ corresponds to the evolution of the phase, which is isomorphic for all of the trajectories, thus $\lambda_\phi = 0$. The remaining exponents, however, can be both positive and negative, and only their sum is restricted to be equal to $-\gamma$. If one of them is positive, the particle's dynamics is chaotic; if both are negative, it is not. In Fig.~\ref{fig:bifurc_m}(b) we plot the maximum Lyapunov exponent $\lambda_\mathrm{max} = \max\{\lambda_x,\ \lambda_v,\ \lambda_\phi\}$ for the  
system in the deterministic limit $\theta=0$ as a function of the mass correction $\Delta m$. On the one hand, in the region of $\Delta m$ where the period-averaged velocity $\mathsf{v}(t)$ is constant, the maximum Lyapunov exponent $\lambda_\mathrm{max} = 0$ and consequently $\lambda_x, \lambda_v < 0$. On the other hand, for higher values of the mass correction, the aperiodic behaviour of $\mathsf{v}(t)$ corresponds to $\lambda_\mathrm{max} > 0$ and therefore the system evolves in a chaotic way. The presence of the constant and aperiodic solutions for $\mathsf{v}(t)$ in the original setup given by Eq. \ref{eq:gle} is thus rooted in the periodic and chaotic character of the particle's dynamics in the  deterministic counterpart of the system.
\begin{figure}[htbp]
\begin{minipage}{0.49\linewidth}
    \centering
    \hspace*{-0.07\linewidth}
    \includegraphics{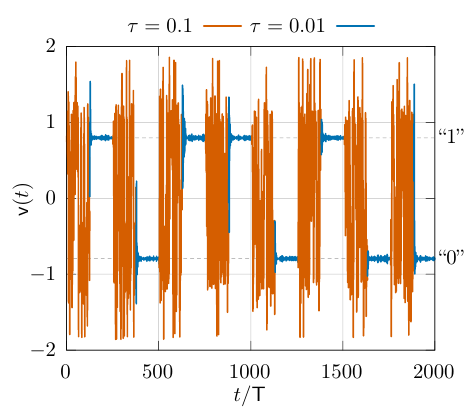}
    \captionof{figure}{Time evolution of the period-averaged velocity $\mathsf{v}(t)$. The memory time $\tau$ switches between values $0.01$ and $0.1$ every $N = 100$ periods of the driving force $\mathsf{T}$. The corresponding bit sequence is ``$10110100$''.}
    \label{fig:traj_switch}
\end{minipage}\hfill
\begin{minipage}{0.49\linewidth}
    \centering
    \vspace*{-24pt}
    \hspace*{-0.02\linewidth}
    \includegraphics{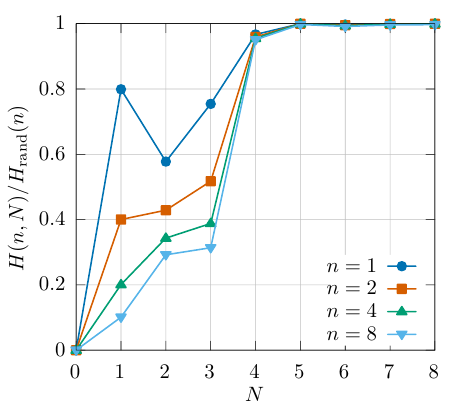}
    \captionof{figure}{Normalized Shannon entropy $H(n, N)/H_\mathrm{rand}(n)$ of the bit segments of length $n$ as a function of the number of chaotic periods $N$.}
    \label{fig:entropy}
\end{minipage}
\end{figure}

We now present how this system can operate as a random bit generator. In Fig.~\ref{fig:traj_switch} we plot the exemplary time evolution of the period-averaged velocity $\mathsf{v}(t)$ of the Brownian particle. Every $N = 100$ periods of the driving force $\mathsf{T}$ the memory time $\tau$ of the thermal bath is switched between $\tau = 0.1$ (chaotic state) and $\tau=0.01$ (bistable state). We adopt the convention that the positive $\mathsf{v}_+ = 0.8$ and negative $\mathsf{v}_- = -0.8$ state is identified with "1" and "0" bit, respectively. In this example the number $\mathsf{N}$ of periods spent in the bistable state is the same as in the chaotic one $\mathsf{N} = N = 100$, however the former is associated only with the rate of bits generation and can be adjusted to the needs without the negative impact on the their randomness. The latter characteristic is related to the time interval of $N$ periods in the chaotic state when the bit is quickly lost and the next value is not correlated with the previous one.


To quantify the randomness of the generated bit values, we calculate the Shannon entropy $H(n, N)$ of the bit sequences of different lengths $n$ generated as in Fig.~\ref{fig:traj_switch} as a function of the number of chaotic periods $N$ \cite{Shannon1948}. To estimate $H(n, N)$, we first replace the trajectory with a sequence of bit values with $\mathsf{v}_+$ corresponding to $1$ and $\mathsf{v}_-$ corresponding to $0$. Then we divide the sequence into segments of length $n$ and calculate the probability $p_i$ of occurrence of each of the $2^n$ possible combinations of $n$ bits. Then the entropy is calculated as
\begin{equation} \label{eq:H}
    H(n, N) = -\sum\limits_{i=1}^{2^n} p_i \log_2(p_i).
\end{equation}
For a completely random sequence all the probabilities $p_i = 1/2^n$ and the entropy equals $H_\mathrm{rand}(n) = n$. The bit sequence can be considered random if the entropy $H(n, N)$ is close to $H_\mathrm{rand}(n)$ for all segment lengths $n$ for which the estimation of $p_i$ is statistically reliable (i.e.~the number of segments is much greater than the number of possible combinations $2^n$). The normalized Shannon entropy $H(n, N)/H_\mathrm{rand}(n)$ calculated for our system is presented in Fig.~\ref{fig:entropy}. Intuitively, the bit sequence obtained in our setup is more random when the chaotic part of the trajectory is longer ($N$ is larger). From Fig.~\ref{fig:entropy} it follows that after $5$ or more chaotic periods the Shannon entropy of the bit sequence is roughly the same as $H_\mathrm{rand}(n)$ for all $n \leq 8$ and as such, it can be considered random.
\begin{figure}[htbp]
    \centering
    \includegraphics{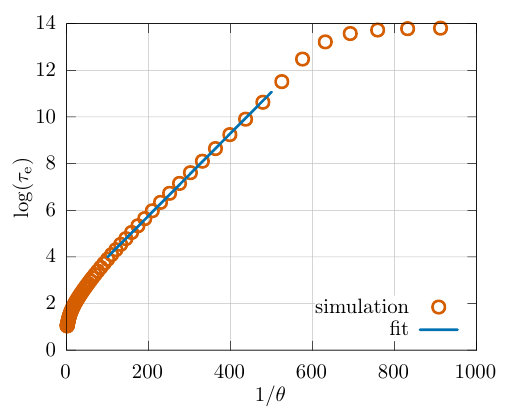}
    \caption{Logarithm of the mean escape time $\tau_\mathrm{e}$ from the attractors corresponding to the bistable period-averaged velocity states as a function of the inverse temperature $1/\theta$. The straight line is fitted to simulation results for the intermediate temperatures $\theta \in [0.002, 0.01]$.}
    \label{fig:kramers}
\end{figure}

Finally, we assess the stability of the constant solutions for $\mathsf{v}(t)$. For $\tau < 0.02$ the particle can be considered to be in a bistable potential in the period-averaged velocity domain. To switch between the potential minima corresponding to the velocities $\mathsf{v}_\pm$, the particle needs to overcome some energy barrier $\Delta E$. The height of the barrier indicates how stable the periodic solutions are and how the mean escape time from each of the minima depends on the temperature of the bath $\theta$. To estimate the height of the barrier $\Delta E$ we invert the Kramers problem and calculate the mean escape time $\tau_\mathrm{e}$ from each of the periodic attractors as a function of the temperature $\theta$ \cite{Kramers1940, Hanggi1990}. The energy barrier can then be estimated by fitting a line to the calculated quantities based on the equation
\begin{equation} \label{eq:kramers}
    \log(\tau_\mathrm{e}) = \Delta E \frac{1}{\theta} + C,
\end{equation}
where $C$ is a constant. In Fig.~\ref{fig:kramers} we present $\log(\tau_\mathrm{e})$ as a function of the inverse temperature $1/\theta$ and the linear fit based on Eq.~\ref{eq:kramers}.

For low temperatures (high $1/\theta$) the mean escape time $\tau_\mathrm{e}$ is comparable or higher than the simulation time $t = 10^6\times\mathsf{T}$, so the estimation of $\log(\tau_\mathrm{e})$ is not reliable and consequently the plot is not a straight line in that region. Furthermore, for high temperatures (low $1/\theta$), the trajectories of the particle are so noisy that the assignment of the particle to one of the potential wells in the period-averaged velocity space is ambiguous. The line is thus fitted to the data for $1/\theta \in [100,\ 500]$. The estimated height of the energy barrier is then
\begin{equation}
    \Delta E = 0.0177 \pm 0.0001.
\end{equation}
This means that in order to minimize the risk of the random switching between the bit values in the bistable state, the temperature should be much lower than $\theta = 0.0177$. This explains the clear separation of the two stable states in Fig.~\ref{fig:bifurc_tau}(a) calculated for $\theta = 10^{-4}$.

\section{Conclusions}
In this article we presented a setup for encoding information in the dynamics of a Brownian particle in a viscoelastic medium. In particular, we considered a system in which the particle can be either in a bistable or chaotic state, depending on the memory time of the surroundings or correlation time of thermal fluctuations. The bit of information can then be encoded in one of the stable states, and the stored data can be erased by changing the memory time and making the particle's dynamics chaotic. First, we showed that the dynamics of the particle can be controlled by changing the memory time of the bath. Moreover, we showed that an approximately equivalent change can be achieved by applying a correction to the particle's mass in a corresponding memoryless setup. The principle of operation of our memory-controlled random bit generator can thus be applied to any other system, in which the change of the viscoelastic properties of the medium, 
leads to the emergence of qualitatively new solutions that can be utilized for storage or erasure of the information. Next, we quantified the randomness of the generated bit values by calculating the Shannon entropy of segments of the generated bit sequence. Finally, we assessed the stability of the information bits depending on the intensity of thermal fluctuations experienced by the particle. Our study provides a general \emph{modus operandi} for designing similar systems for storing and processing information on a microscopic scale.

\funding{This work was supported by the Grant NCN No. 2024/54/E/ST3/00257 (JS).}

\roles{
MW: conceptualization, data curation, formal analysis, investigation, software, validation, visualization, writing -- original draft\\
\noindent JS: conceptualization, formal analysis, funding acquisition, methodology, resources, software, supervision, writing -- review \& editing
}

\data{The data cannot be made publicly available upon publication because they are not available in a format that is sufficiently accessible or reusable by other researchers. The data that support the findings of this study are available upon reasonable request from the authors.}


\setlength\bibitemsep{0.15\baselineskip}
\printbibliography

\end{document}